\documentclass[aps,amsfonts,pre,twocolumn,superscriptaddress,showpacs,eqsecnum]{revtex4-1}
\usepackage{epsfig,amsmath,amssymb,bm,epsf,graphics,psfrag,verbatim,subfigure}

\usepackage[normalem]{ulem}
\usepackage{float}
\usepackage[usenames]{color}
\usepackage{graphicx}
\usepackage{ulem}
\usepackage{amsfonts}
\usepackage{amsmath}
\usepackage{natbib}
\usepackage{epsfig,amsmath,amssymb,bm,epsf,graphics,psfrag,verbatim,subfigure,inputenc}

\def\kNNN{\kappa}
\def\g4{g}
\def\dg{\Lambda} 
\def\uf{\vec{u}'} 
\def\st0{\epsilon^0} 
\def\kr{\tau} 
\def\gr{\lambda} 
\newcommand{\qv}{\mathbf{q}}
\newcommand{\gn}{n}
\newcommand{\nz}{K}
\newcommand{\tT}{\tilde{T}}

\def\st{\epsilon}
\def\stm{\epsilon^{0}}

\begin{document}

\title{Mechanical instability at finite temperature }

\author{Xiaoming Mao}
\affiliation{Department of Physics, University of Michigan, Ann
Arbor, MI 48109, USA }

\author{Anton Souslov}
\affiliation{School of Physics, Georgia Institute of Technology, Atlanta, Georgia 30332, USA}

\author{Carlos I. Mendoza}
\affiliation{Instituto de Investigaciones en Materiales, Universidad Nacional Aut{\'o}noma de M{\'e}xico, Apdo. Postal 70-360, 04510 M{\'e}xico, D.F., Mexico}

\author{T. C. Lubensky}
\affiliation{Department of Physics and Astronomy, University of
Pennsylvania, Philadelphia, PA 19104, USA }

\date{\today}
\begin{abstract}
Many physical systems including lattices near structural
phase transitions, glasses, jammed solids, and bio-polymer gels
have coordination numbers that place them at the edge of
mechanical instability. Their properties are determined by an
interplay between soft mechanical modes and thermal
fluctuations. In this paper we investigate a simple
square-lattice model with a $\phi^4$ potential between
next-nearest-neighbor sites whose quadratic coefficient
$\kappa$ can be tuned from positive negative. We show that its
zero-temperature ground state for $\kappa <0$ is highly
degenerate, and we use analytical techniques and simulation to
explore its finite temperature properties. We show that a
unique rhombic ground state is entropically favored at nonzero
temperature at $\kappa <0$ and that the existence of a
subextensive number of ``floppy" modes whose frequencies vanish
at $\kappa = 0$ leads to singular contributions to the free
energy that render the square-to-rhombic transition first order
and lead to power-law behavior of the shear modulus as a
function of temperature. We expect our study to provide a
general framework for the study of finite-temperature
mechanical and phase behavior of other systems  with a large
number of floppy modes.

\end{abstract}

\maketitle

\section{Introduction}
Crystalline solids can undergo structural phase transitions in
which there is a spontaneous change in the shape or internal
geometry of their unit cells
\cite{FolkSch1976,Cowley1980,Bruce1980,Fujimoto2005}. These
transitions are signalled by the softening of certain elastic
moduli or of phonon modes at a discrete set of points in the
Brillouin zone. Lattices with coordination number $z=z_c=2d$ in
$d$ spatial dimensions, which we will call Maxwell lattices
\cite{Maxwell1864}\footnote{The term \emph{isostatic} is
often incorrectly used to describe any system with $z=z_c$.
Finite Isostatic lattices have $z=z_c^N=z_c -d(d+1)/N$ and no
states of self stress - See~\cite{Calladine1978} for example.
There is no universally accepted definition of \emph{isostatic}
in lattices with periodic boundary conditions, but one such as
the square lattice with many states of self-stress is surely
not isostatic.}, exist at the edge of mechanical instability,
and they are critical to the understanding of systems as
diverse as engineering structures
\cite{Heyman1999,Kassimali2005}, diluted lattices near the
rigidity threshold \cite{Feng1984,FengLob1984,Jacobs1995},
jammed systems \cite{Liu1998,Wyart2005a,LiuNag2010a},
biopolymer networks
\cite{Elson1988,Kasza2007,Alberts2008,Janmey1990,BroederszMac2014},
and network glasses \cite{Phillips1981,Thorpe1983}. Hypercubic
lattices in $d$ dimensions and the kagome lattice and its
generalization to higher dimensions with nearest-neighbor (NN)
Hookean springs of spring constant $k$ are a special type of
Maxwell lattice whose phonon spectra have harmonic-level zero
modes not at a discrete set of points but at all $N^{(d-1)/d}$
points on $(d-1)$-dimensional hyperplanes oriented along
symmetry directions and passing through the origin
\cite{Souslov2009}. A question that arises naturally is whether
these lattices can be viewed as critical lattices at the
boundary between phases of different symmetry and, if so, what
is the nature of the two phases and the phase transition
between them.

\begin{figure}
	\centering
    \includegraphics[width=0.43\textwidth]{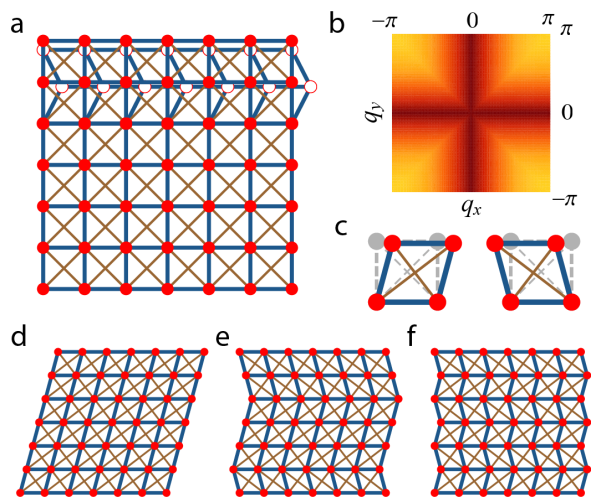}
	\caption{(a) The square lattice model with NN (blue thick) bonds and NNN (brown thin)
    bonds.  White disks showing a shift of the second row is one example of a floppy
    modes of the lattice with no NNN bond. (b) Density plot of the phonon spectrum
    of the NN square lattice showing lines of zero modes (darker color corresponds to lower frequency).
    (c) The $T=0$ ground states of a plaquette when $\kNNN < 0$, with the undeformed
    reference state shown in gray.  (d-f) Examples of $T=0$ ground states of the
    whole lattice when $\kNNN < 0$. (d) shows the uniformly sheared rhombic lattice,
    which we show to be the preferred configuration at small $T$ in the thermodynamic
    limit. (e) is a randomly zigzagging configuration, and (f) is the ordered maximally
    zigzagging configuration, which has a unit cell consisting of two particles.
	}
\label{Fig:Model}
\end{figure}

Here we introduce and study, both analytically and with
Monte-Carlo simulations, a square-lattice model (easily
generalized to higher dimensions) in which
next-nearest-neighbors (NNNs) are connected via an anharmonic
potential consisting of a harmonic term with a
 spring constant $\kNNN$ tuned from
positive to negative and a quartic stabilizing term. When
$\kNNN>0$, the square lattice is stable even at zero
temperature. When $\kNNN=0$, NNN springs contribute only at
anharmonic order, and the harmonic phonon spectrum is identical
to that of the NN-lattice. When $\kNNN<0$, the NNN potential
has two minima, and the ground state of an individual plaquette
is a rhombus that can have any orientation (Fig.~\ref{Fig:Model}c). 
Plaquettes in the same row (or column), however, are
constrained to have the same orientation, but plaquettes in
adjacent rows can either tilt in the same direction or in the
mirror-image direction as shown in Figs.~\ref{Fig:Model}d-f,
leading to $2\times 2^{\sqrt{N}}$ equivalent ground states and
a subextensive but divergent entropy of order $\sqrt{N}\ln2$.
The properties of this model, including the subextensive
entropy at zero temperature, are very similar to those of
colloidal particles confined to a low-height cells
\cite{Pieranski1983,Han2008} and to the anti-ferromagnetic
Ising model on a deformable triangular lattice
\cite{Shokef2011}. In addition, the scaling of the shear
modulus near the zero-temperature critical point is analogous
to that observed in finite-temperature simulations of randomly
diluted lattices near the rigidity-percolation threshold
\cite{DennisonMac2013} and to finite-temperature scaling near
the jamming transition \cite{IkedaBir2013}, suggesting that
generalizations of our model and approach may provide useful
insight into the thermal properties of other systems near the
Maxwell rigidity limit.

\section{Results}
\begin{figure}
	\centering
		\includegraphics[width=.42\textwidth]{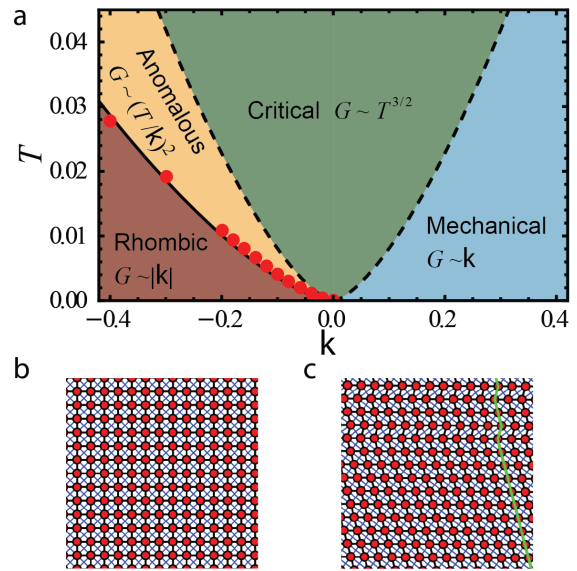}
	\caption{(a) An example of the phase diagram of the model square lattice at $k=1$,
    $\g4=10$, $a=1$.  The black solid line (the red dots) shows the boundary obtained from analytic theory (Monte Carlo simulation) between the square phase on the right and the rhombic phase on the left.  The square phase is stabilized by thermal fluctuations even for $\kNNN<0$ and the shear modulus $G$ of the square lattice exhibit different scaling regimes (separated by black dashed lines) determined by Eq.~\eqref{EQ:hsscaling}.  (b) and (c) show Monte Carlo snapshots of the square and rhombic phases respectively.  A small number of zigzags exist in the rhombic phase Monte Carlo snapshots, resulting from finite size effects as discussed in Sec.~\ref{sec:model}.
	}
	\label{FIG:PD}
\end{figure}

Strong fluctuations arising from the large number of zero modes
lead to interesting physics at $T>0$ in this square lattice model. We show  the following:
\begin{itemize}
\item Among all the equal-energy zigzagging configurations
    at $\kNNN<0$, the uniformly sheared rhombic lattice,
    shown in Fig.~\ref{Fig:Model}d, has the lowest free
    energy: the ground-state degeneracy is broken by
    thermal fluctuations through an order-by-disorder
    effect~\cite{Villain1980,Shender1982,Henley1987,Henley1989,
    Chubukov1992,Reimers1993, Bergman2007, Shokef2011}.
\item Thermal fluctuations lead to a negative coefficient
    of thermal expansion and corrections to the shear
    rigidity
    \cite{RubinsteinBas1992,Barriere1995,Plischke1998,TessierDis2003}
    that enable the square lattice state to remain
    thermodynamically stable in a region of the phase
    diagram at $\kNNN\leq0$ (Fig.~\ref{FIG:PD}).
\item Fluctuations drive the transition from the rhombic to
    the square phase first order and lead to the phase
    diagram shown in Fig. \ref{FIG:PD} in which the
    temperature of the transition approaches zero as
    $\kappa \rightarrow 0^-$.
\item The low-$T$  shear modulus $G$ (Fig.~\ref{FIG:PD})
    in both phases is proportional to $|\kappa|$ at low temperature, and there
    is a critical regime with $T > \text{const.}
    \times\kappa^{3/2}$ in which $G\sim T^{3/2}$.
    In addition, there is a region in the square phase (mostly metastable with
    respect to the rhombic phase)  with $T  < \text{const.} \times
    |\kappa|^{3/2}$ in which $G \sim (T/|\kappa|)^2$.
    This behavior near the $T=0$ critical point is analogous to
    that found in the randomly diluted triangular lattice
    \cite{DennisonMac2013} near the central-force rigidity
    threshold.  Interestingly, the critical regime in our
    model is fundamentally a consequence of nonlinearity as
    is the case for dynamical scaling near the jamming
    transition \cite{IkedaBir2013}.
\end{itemize}

These predictions are supported by our Monte-Carlo simulations
and by direct calculations of entropic contributions to the
free energy phonon fluctuations in different arrangements of
kinks.

\section{Discussion}
Our model of the square-to-rhomic transition is very similar to
a model, studied by Brazovskii \cite{Brazovskii1975}, for the
transition from an isotropic fluid to a crystal and later
applied to the Rayleigh-B\'{e}nard instability
\cite{SwiftHoh1977} and to the nematic-to-smectic-$C$
transitions in liquid crystals \cite{ChenLub1976,Swift1976}. In
all of these systems, there is a subextensive but infinite
manifold of zero modes [a $(d-1=2)$-dimensional spherical shell
in the Brazovskii case, a $(d-1=1)$-dimensional circle in the
Rayleigh-B\'{e}nard case, and two $(d-2=1)$-dimensional circles
in the liquid crystal case] in the disordered phase leading to
a singular contribution to the free energy. We use the
Brazovskii theory to calculate the temperature of the
first-order square-to-rhomic transition as a function of
$\kNNN<0$ and negative thermal expansion in the square phase.

In applying the Brazovskii approach to our problem, we develop
an expansion of the  free energy that maintains rotational
invariance of elastic distortions in the target space. Previous
treatments \cite{FolkSch1976} of structural transitions tend to
mix up nonlinear terms arising from nonlinearities in the
strain tensor required to ensure rotational invariance and
nonlinear terms in the elastic potential itself.  This advance
should be useful for the calculation of renormalized free
energies and critical exponents in standard structural phase
transitions.

The number and nature of zero modes of the critical NN square
lattice has a direct impact on the properties of the lattices
with $\kNNN>0$ and $\kNNN<0$, and it is instructive to review
their origin.  A powerful index theorem \cite{Calladine1978}
relates the number of zero modes of a frame consisting of $N$
points and $N_B \equiv \frac{1}{2} z N$ bonds, where $z$ is the
average coordination number, via the relation
\begin{equation}
N_0 = d N - N_B + S ,
\end{equation}
where $S$ is the number of independent states of self-stress in
which bonds are under tension or compression and in which the
net force on each point is zero. In his seminal $1864$ paper
\cite{Maxwell1864}, Maxwell considered the case with $S=0$ that
yields the Maxwell relation for the critical coordination
number at which $N_0$ is equal to the number $\gn(d)$ ($=d$ for
periodic and $d(d+1)/2$ for free boundary conditions) of zero
modes of rigid translation and rotation:
\begin{equation}
z_c^N = 2 d - \frac{2 \gn(d)}{N} .
\end{equation}
In the limit of large $N$, $z_c^N \rightarrow z_c^\infty = 2
d$.

There are many small unit cell periodic Maxwell lattices with
$N_0 = S$. The NN square and kagome lattices in two dimensions
and, the cubic and pyrochlore lattices in three dimensions are
special examples of these lattices that have sample-spanning
straight lines of bonds that support states of self-stress
under periodic boundary conditions.  They, therefore, have of
order $N^{(d-1)/d}$ states of self-stress and the same number
of zero modes, which are indicators of buckling instabilities
of the lines when subjected to compression. Geometrical
distortions of theses lattices that remove straight lines, as
is the case with the twisted kagome lattice \cite{Sun2012},
remove states of self-stress and associated zero modes.  When
subjected to free rather than periodic boundary conditions,
these distorted lattices have a deficiency of order
$N^{(d-1)/d}$ bonds and as a result the same number of surface
zero modes (there are no bulk zero mode other than the trivial
ones of uniform translation), which can have a topological
character \cite{Kane2014} or be described in the
long-wavelength limit by a conformal field theory
\cite{Sun2012}. Unlike the infinitesimal zero modes of
hypercubic lattices, those of the kagome and pyrochlore do not
translate into finite zero modes of the lattices when
finite sections are cut from a lattice under periodic boundary
conditions. Thus, it is not yet clear whether the ground state
of the latter lattices will are highly degenerate or not.
Nevertheless, Brasovskii theory should provide a sound
description of thermal properties of these lattices in the
vicinity of the point $T=0$, $\kappa=0$.

\section{Model and order-by-disorder in the low-symmetry phase}
\label{sec:model}
The model we consider is a square lattice with two different
types of springs -- those connecting nearest neighbors and
those connecting next-nearest neighbors, as shown in
Figure~\ref{Fig:Model}(a). The NN springs
are Hookian, with potential
\begin{equation}
	V_{\textrm{NN}}\left(x\right) = \frac{k}{2} x^2
\end{equation}
where $k > 0$. The NNN springs are introduced with an
anharmonic potential
\begin{equation}
	V_{\textrm{NNN}}\left(x\right) = \frac{\kNNN}{2} x^2 + \frac{\g4}{4!} x^4,
\end{equation}
where $\kNNN$ can be either positive or negative and $\g4$,
introduced for stability, is always positive. The Hamiltonian
of the whole lattice is thus,
\begin{align}\label{EQ:Hami}
	H =& \sum_{\langle i,j\rangle\in\textrm{NN}}
	V_{\textrm{NN}}\left(\vert \vec{R}_i - \vec{R}_j \vert- a\right) \nonumber\\
	&+ \sum_{\langle i,j\rangle\in\textrm{NNN}}
	V_{\textrm{NNN}}\left(\vert \vec{R}_i - \vec{R}_j \vert-
\sqrt{2}a\right),
\end{align}
where $\vec{R}_i$ is the positions of the node $i$ and $a$ is
the lattice constant. In what follows, we will use the reduced
variables
\begin{equation}
\kr\equiv \kNNN/k  \qquad \text{and} \qquad \gr\equiv \g4 a^2/k
\end{equation}
to measure the strength of couplings in $V_{\textrm{NNN}}$.

For $\kNNN > 0$, $V_{\textrm{NNN}}\left(x\right)$ has a unique
minimum at $x = 0$, and the ground state of $H$ is the square
lattice with lattice spacing $a$. All elastic moduli of this
state are nonzero, and it is stable with respect to thermal
fluctuations, though, as we shall see, it does undergo thermal
contraction at nonzero temperature.  When $\kNNN<0$,
$V_{\textrm{NNN}}\left(x\right)$ has two minima at $x = \pm
\sqrt{6 \kNNN/\g4}$, corresponding to stretch and compression,
respectively.  This change in length of NNN springs is resisted
by the NN springs, and in minimum energy configurations one NNN
bond in each plaquette will stretch and the other will
contract.  The alternative of having both stretch or contract
would cost too much NN energy. A reasonable assumption, which
is checked by our direct calculation, is that the stretching
and contraction will occur symmetrically about the center so
that the resulting equilibrium shape is a rhombus rather than a
more general quadrilateral (see Supplementary Information Sec. I). The shape of a rhombus is uniquely
specified by the lengths $d_1$ and $d_2$ of its diagonals
(which are perpendicular to each other), whose equilibrium
values are obtained by minimizing the sum over plaquettes of
\begin{eqnarray}
V_{\textrm{PL}}& = &2 V_{\textrm{NN}} (\frac{1}{2}\sqrt{d_1^2 +
d_2^2} - a) + \nonumber\\
& & V_{\textrm{NNN}} (d_1 - \sqrt{2}a) +
V_{\textrm{NNN}} (d_2 - \sqrt{2}a) .
\end{eqnarray}
The grounds state of the entire lattice when $\kNNN <0$ must
correspond to a tiling of the plane by identical rhombi each of
whose vertices are four-fold coordinated. It is clear that
zigzag arrangements of rows (or columns) of rhombi in which
adjacent rows tilt in either the same or opposite directions
constitute a set of ground states.  A derivation showing that
this is the complete set can be found in
Ref.~\cite{Shokef2011}, which considered packing of isosceles
triangles, which make up half of each rhombus. The ground state
energy per site, $\epsilon_0$, is simply $V_{\textrm{PL}}$
evaluated at the equilibrium values of $d_1$ and $d_2$.

Each ground-state configuration of a system with $N_x$ vertical
columns and $N_y$ horizontal rows has $\nz=0, \cdots , N_y$
horizontal zigzags or $\nz=0, \cdots , N_x$ vertical zigzags.
Thus, the ground state entropy diverges in the thermodynamic
limit, though it is sub-extensive and proportional to $N_x +
N_y$ in a  system of $N = N_x N_y$ particles. Such ground state
configurations are found in other systems, most notably the
zigzagging phases seen in suspensions of confined colloidal
particles~\cite{Han2008,Pieranski1983}. The confined colloidal
system has a phase diagram that depends only on the planar
density and the height of confinement of the colloids. For
sufficiently large heights, the colloids form a phase of two
stacked square lattices. In a neighboring region of the phase
diagram, explored in simulations of
Refs.~\cite{Schmidt1996,Schmidt1997}, this square lattice
symmetry is broken through a weakly discontinuous transition
and a rhombic phase is observed. This region of the phase
diagram of confined colloids thus provides a physical
realization of the Hamiltonian~(\ref{EQ:Hami}).

\begin{figure}
	\centering
     \includegraphics{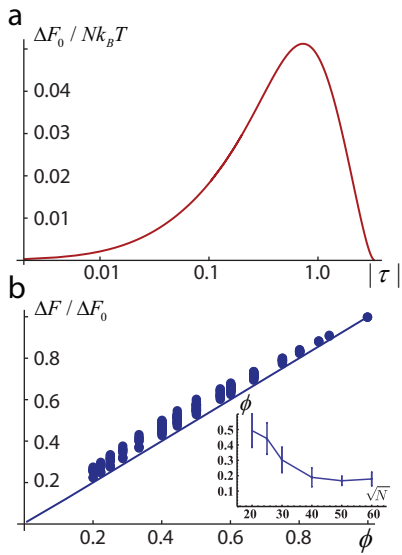}
\caption{ The phonon contribution to the free energy for
various zigzagging configurations. (a) shows the free energy
difference $\Delta w_0$ between the maximally zigzagging
configuration (Fig.~\ref{Fig:Model}e) and the uniformly sheared
square lattice configuration (Fig.~\ref{Fig:Model}c). For
sufficiently large $\tau$, the lattice collapses onto a line,
at which point the free energy difference goes to zero. (b)
shows the free energy of the phonons as a ratio $\Delta F /
\Delta F_0$ for $\tau = -0.02$, $-0.1$ and $-1$ ($\lambda =
10$), where $\Delta F$ and $\Delta F_0$ are defined in
Eq.~(\ref{Eq:DeltaF}) and $\Delta F$ is evaluated for all
possible configurations with unit cells with at most $10$ sites
as a function of the zigzag fraction $\phi = \nz/N_x$. The
lattice without zigzags is entropically favored for any any
value of $\kr$ in our calculations.  $\Delta F/\Delta F_0$ is
well approximated by the line $\phi$, which corresponds to
non-interacting zigzags. This interaction results in the
dispersion of values of  $\Delta F/\Delta F_0$.  
The inset of (b) shows Monte Carlo simulation results 
for the zigzagging fraction $\phi$ plotted against the linear 
system size $\sqrt{N}$.}
\label{fig:o-by-diso}
\end{figure}

At low but nonzero temperatures, the degeneracy of the ground
state is broken by thermal fluctuations through the
order-by-disorder mechanism~\cite{Villain1980, Henley1987,
Henley1989, Chubukov1992, Reimers1993, Bergman2007,
Shokef2011}. This splitting of degeneracy due to small phonon
fluctuations around the ground state may be calculated using
the dynamical matrix in the harmonic approximation to the
Hamiltonian~(\ref{EQ:Hami}). For each ground state
configuration $\vec{R}_i$, we write the deformation as $\vec{R}_i \to \vec{R}_i
+ \vec{u}_i$, and expand to quadratic order in $\vec{u}$, $H =
\frac{1}{2} \sum_{\langle i j\rangle}\vec{u}_i \mathbf{D}_{ij}
\vec{u}_j$. The Fourier transform of $\mathbf{D}_{ij}$,
$\mathbf{D}_q$, is block-diagonal and the phonon free energy,
which is purely entropic, for that configuration is
\begin{equation}
F_p(N_x,N_y,\nz) = \frac{1}{2} k_B T \sum_\qv \ln \det \mathbf{D}_\qv
\equiv N k_B T w_p,
\label{eq:sp}
\end{equation}
where $w_p$ is the free-energy per site in units of $k_B T$. In
general, $F_p$ depends not only on $\nz$, but also on the
particular sequence of zigzags.  We numerically calculated this
free energy for all periodically zigzagged configurations with
up to 10 sites per unit cell.  We found that the
lowest-free-energy state is the uniformly sheared state
[Fig.~\ref{Fig:Model}(c)] with $\nz = 0$, and the
highest-free-energy state is the maximally zigzagged sate
[Fig.~\ref{Fig:Model}(e)] with $\nz = N_y$ and two sites per
unit cell. Both of these energies are extensive in the number
of sites $N$, and we define
\begin{eqnarray}
\Delta F_0 (N_x,N_y)& = & F_p(N_x,N_y,N_y) - F_p(N_x,N_y, 0 )\nonumber \\
&\equiv & N_x N_y k_b T \Delta w_0 \label{Eq:DeltaF}\\
\Delta F(N_x,N_y,\nz) & = & F_p(N_x,N_y,\nz) - F_p(N_x,N_y,0) . \nonumber
\end{eqnarray}
Figure~\ref{fig:o-by-diso}(a) displays our calculation of $\Delta w_0
(\tau)$, which vanishes, as expected, at $\tau = 0$ and also at
large $\tau$ at which the rhombus collapses to a line.  Figure
\ref{fig:o-by-diso}(b) plots $\Delta F/\Delta F_0$ as a
function of $\phi = \nz/N_y$ for different values of $\tau$. By
construction, this function must vanish at $\phi = 0$ and be
equal to one at $\phi = 1$.  All of the points lie
approximately on a straight line of slope one.  
Thus, we can
approximate $F_p(N_x,N_y,\nz)$ by
\begin{equation}
F_p(N_x,N_y,\nz) = F_p(N_x,N_y,0)+ N_x \nz k_B T \Delta w_0 .
\label{eq:Fp}
\end{equation}
Note that for each $\nz$, this energy is extensive in $N=N_x
N_y$ as long as $\phi \neq 0$.

These calculations were carried out in the thermodynamics
limit, $N \rightarrow 0$, in which the sum over $\qv$ is
replaced by the continuum limit by an integral. In order to
compare these results with the Monte Carlo results of the next
section, it is necessary to study finite size effects. The
first observation is that for any finite $N_x$, the system will
be effectively one-dimensional for a sufficiently large $N_y$,
and as a result, we would expect the number of zigzags to
fluctuate. To proceed, we continue to use the continuum limit
to evaluate Eq.~(\ref{eq:sp}), and we use Eq.~(\ref{eq:Fp}) for
$F_p(N_x,N_y,\nz)$. The partition function for this energy is
\begin{eqnarray}
Z&= &\sum_{\nz=0}^{N_y} e^{-\beta[F_p(N_x,N_y,0)+E_0(N_x,N_y)]}
\binom{N_y}{\nz}e^{-N_x \nz \Delta w_0} \nonumber\\
& = & e^{-\beta[F_p(N_x,N_y,0)+E_0(N_x,N_y)]}
\left(1+ e^{-N_x \Delta w_0}\right)^{N_y}
\end{eqnarray}
where $E_0(N_x,N_y)$ denotes the potential energy which is independent of $K$, and the full free energy is
\begin{align}
& F(N_x,N_y,T) = -T \ln Z \nonumber \\
& = N_x N_y (f_0 + e_0) - N_y k_B T \ln (1 + e^{- N_x \Delta w_0}) ,
\end{align}
where $e_0$ and $f_0$ are the potential energy and phonon free energy per site of the uniformly sheared state.  
Thus, as expected, when $N_x \Delta w_0 \gg 1$, zigzag
configurations make only a very small, subextensive
contribution to the free energy.  On the other hand, in the
opposite limit, they make an extensive contribution of $N_x N_y
k_B T \Delta w_0$ to the energy.  Therefore, at a given $\tau$, the zigzag configurations are favored when the system is small, and in thermodynamic limit, the rhombic configuration is always favored.  Our Monte Carlo simulation verified this  [see inset of Fig.~\ref{fig:o-by-diso}(b)].


\section{Simulation}
We simulate the system using a Monte Carlo (MC) algorithm
inside a periodic box whose shape and size are allowed to
change in order to maintain zero pressure. In this version of
the Metropolis algorithm, also used in
Refs.~\cite{Shokef2009,Shokef2011}, for each MC step a particle
is picked at random and a random trial displacement is
performed. The trial displacement is initially uniformly
distributed within a radius of $0.1 a$, but throughout the
simulation the radius is adjusted to keep the
acceptance probability between $0.35$ and $0.45$. Given the
initial configuration energy $E_i$ and the trial configuration
energy $E_j$, the trial configuration is accepted with
probability $[1 + \exp(E_i - E_j)/T]^{-1}$, i.e., Glauber
dynamics is used. After initializing
the system using the square lattice configuration with lattice
constant $a$, the simulation is first run at a high temperature
and is then annealed to the final low temperature. For each
intermediate temperature, an equilibration cycle in a
sample of $N$ sites consists of at least $N^{4}\times10^{2}$
MC steps. To accommodate areal and shear distortions in the
different phases we encounter, the simulation box area and
shape are changed using a similar acceptance algorithm, with
the trial deformation adjusted to keep the acceptance
probability between $0.35$ and $0.45$, such that the simulation
box retains the shape of a parallelogram~\cite{Frenkel2001}. The simulation is
thus performed at zero pressure, and a range of temperatures
measured in units of $k a^2$ for up to $N=3600$ sites.

We use these simulations to investigate the phase diagram
corresponding to the
Hamiltonian~(\ref{EQ:Hami})~\cite{Binder1981a} and to
investigate the properties of the phases we encounter, such as
ground state degeneracy, order-by-disorder and negative thermal
expansion. As all simulations involve a finite lattice and are
run for a finite time, we took care to make sure that the
system is sufficiently large to capture the thermodynamic
behavior and that the simulation time is sufficiently long for
the system to relax to equilibrium. To capture the
subtlety of the order by disorder effect for a finite system,
we simulated the model for a range of sizes and times and
calculated the average fraction of zigzags $n$ in equilibrium
[inset of Fig.~\ref{fig:o-by-diso}(b)]. While for small systems, $\phi
\approx \frac{1}{2}$, as in a disordered zigzagging
configuration, for large systems, $\phi$ approaches $0$,
suggesting that the system prefers the configuration of a
uniformly sheared square lattice. Thus, we find good agreement
with theoretical results from Sec.~\ref{sec:model}.

\begin{figure}
	\centering
     \includegraphics{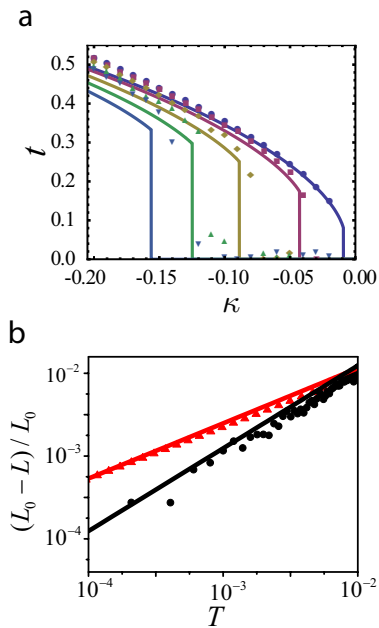}
	\caption{(a) Order parameter $t$ calculated from theory (lines) and simulation (data points),
    at $\gr=10$, and from right to left, $T/(ka^2)=0.0001, 0.001, 0.003, 0.005, 0.007$.
    (b)Negative thermal expansion.  Shown in the figure are normalized size change $(L_0-L)/L_0$ as
    a function of $T$ at $\kappa=0,g=10$ (red triangles: MC data; red upper line: theory),
    and at $\kappa=0.1, g=0$ (black circles: MC data; black lower line: theory).}
	\label{fig:OPNTE}
\end{figure}

We calculated the shape of the phase boundary shown in
Fig.~\ref{FIG:PD}, the behavior of the order parameter across
the phase boundary shown in Fig.~\ref{fig:OPNTE}a, and the
negative of the thermal expansion coefficient $(L-L_0)/L_0$
shown in Fig.~\ref{fig:OPNTE}(b), where $L$ is the length at
$T>0$ and $L_0$ that at $T=0$. The phase boundary is obtained
by calculating the heat capacity of the system as a function of
temperature at fixed $\lambda$ and $\tau$. The location of the
peak of the heat capacity corresponds to the location of the
phase transition in the thermodynamic limit, and in our
simulations, the locations of the peak converge to the values
seen in Fig.~\ref{FIG:PD}. The order parameter values are
calculated locally, i.e.,  $t$ is calculated for each
plaquette and each configuration via the angle between the two
adjacent nearest-neighbor bonds and then averaged over all
plaquettes in the system and over all 100
configurations. In this approach, $t$ is independent of the
particular zigzag configuration, and its evaluation does not
exhibit the long relaxation process to the uniformly sheared
square lattice. The behavior of the order parameter in
Fig.~\ref{fig:OPNTE} is consistent with a weakly-discontinuous
transition.  

\section{Analytic theory and the Phase Diagram}\label{sec:theory}

The special feature of our model is its large but subextensive
number of soft modes living on the $q_x$ and $q_y$ axes in the
first Brillouin zone,  as shown in Fig.~\ref{Fig:Model}(b) in the
limit $\kNNN\to 0$)~\cite{Souslov2009,Mao2010}.  As we discuss
below, these floppy modes provide a divergent fluctuation
correction to the rigidity of the square lattice and render the
transition from the square to the rhombic phase first order.
Thus, this model is analogous to the one introduced by
Brazovskii \cite{Brazovskii1975} for a liquid-solid transition
in which mode frequencies of the form
$\omega=\Delta+(q-q_c)^2/m$ vanish on a $(d-1)$-dimensional
hypersphere when $\Delta \rightarrow 0$ and contribute terms to
the free energy singular in $\Delta$.

To study the square-to-rhombic transition, we take the $T=0$
square lattice, with site $i$ at position $\vec{R}_{0,i}$, as
the reference state.  We then represent positions in the
distorted lattices as the sum of a part arising from uniform
strain characterized by a deformation tensor $\dg$ and
deviations $\uf_{i}$ from that uniform strain:
\begin{align}
\vec{R}_{i} = \mathbf{\dg} \cdot \vec{R}_{0,i} + \uf_{i} ,
\label{eq:R0i}
\end{align}
The deviations $\uf_{i}$ are constrained to satisfy periodic
boundary conditions, and their average $\langle \uf_{i}
\rangle$ is constrained to be zero. The former condition
ensures that the sum over all bond stretches arising from these
deviations vanishes for every configuration. Without loss of
generality, we take $\Lambda_{yx}$ to be zero, leaving three
independent parameters to parameterize the three independent
strains.
As we detail in Supplementary Information Sec. IV, the strain parameter
characterizing pure shear with axes along the $x$ and $y$ axes
of the reference lattice is of order $t^2$ and can be
ignored near $\tau = 0$, and we set
\begin{align}\label{EQ:Lambda}
	\mathbf{\dg} = \left( \begin{array}{cc}
	1+s & t \\
	0 & 1+s
	\end{array} \right) .
\end{align}
Note that $\dg$ is invertible even though it is not symmetric.
$t$ is the order parameter that distinguishes the rhombic phase
from the square phase.  Thermal fluctuations lead to $s<0$ in
both phases.

Expanding the Hamiltonian [Eq.~(\ref{EQ:Hami})] in to second
order $\uf_{i}$ about the homogeneously deformed state $\langle
\vec{R}_{i} \rangle=\mathbf{\dg} \cdot \vec{R}_{0,i}$, we
obtain
\begin{align}\label{EQ:HExpa}
	H = H_0(\mathbf{\dg}) + \frac{1}{V} \sum_{q} \uf_{q} \cdot \mathbf{D}_q(\mathbf{\dg}) \cdot \uf_{-q}
	+  O((\uf)^3) ,
\end{align}
where $H_0$ is the energy of the uniformly deformed state, and
\begin{align}\label{EQ:DML}
	\mathbf{D}_q(\mathbf{\dg}) = v_q(\mathbf{\dg}^T \mathbf{\dg}) \mathbf{I}
	+ \mathbf{\dg}\cdot \mathbf{M}_q(\mathbf{\dg}^T \mathbf{\dg}) \cdot \mathbf{\dg}^{T}
\end{align}
is the $d\times d$ dimensional ($d=2$ being spatial dimension)
dynamical matrix  with scalar $v_q$ and second rank tensor
$\mathbf{M}_q$ determined by the potentials. There is no term
linear in $\uf_{i}$ in Eq.~(\ref{EQ:HExpa}) because of the
periodicity constraint (see Supplementary Information Sec. II).

Integrating out the fluctuations $\uf$ from the
Hamiltonian~\eqref{EQ:HExpa}, we obtain the free energy of the
deformed state~\footnote{This form is similar to
Eq.~\ref{eq:sp}, except that here we expand around uniformly
deformed lattice of continuously varying $\dg$, rather than
about a zigzagging state.}
\begin{align}
\label{eq:f}
	F(\mathbf{\dg}) = H_0(\mathbf{\dg}^T \mathbf{\dg}) +
\frac{T}{2}\ln\textrm{Det}\tilde{\mathbf{D}}(\mathbf{\dg}) ,
\end{align}
where $\tilde{\mathbf{D}}=v_q + M_q \mathbf{\dg}^T
\mathbf{\dg}$ depends only on $\mathbf{\dg}^T \mathbf{\dg}=1+2
\mathbf{\st0}$, where $\mathbf{\st0}$ is the full nonlinear
strain.  Thus the one-loop free energy of Eq.~(\ref{eq:f}) is a
function of the nonlinear, rather than the linear strain, so
that rotational invariance in the target space is guaranteed
and there is a clean distinction between nonlinear terms in
linearized deformations arising from nonlinearities in
$\mathbf{\st0}$ in from nonlinear terms in the expansion in
powers of $\mathbf{\st0}$.

To analyze the transition between the square and the rhombic
phases at low temperature we expand $F$ as a series in
$\mathbf{\st0}$, by expanding the transformed dynamical matrix
as $\tilde{\mathbf{D}} = \mathbf{D}_0
+\mathbf{A}(\mathbf{\st0})$, where $ \mathbf{D}_0 =
\mathbf{D}\vert_{\mathbf{\st0} = 0}$ is the dynamical matrix of
the undeformed state. The free energy is then
\begin{align}\label{eq:Fexpansion}
	F(\mathbf{\st0}) = H_0(\mathbf{\st0}) + \frac{T}{2}\textrm{Tr} \ln
	 \mathbf{D}_0 \left\lbrack \mathbf{I}+\mathbf{G}_0 \cdot \mathbf{A}(\mathbf{\st0})\right\rbrack
    \equiv V f(\mathbf{\st0}) ,
\end{align}
where $\mathbf{G}_0 \equiv \mathbf{D}_0 ^{-1}$ is the phonon
Green's function in the undeformed state, $V=Na^2$ and $f$ is
the free energy density. The expansion of $F$ at small
$\mathbf{\st0}$ follows from this.

Close to the transition, $F$ is dominated by fluctuations
coming from the floppy modes as we discussed above.  As $\kNNN\to
0$, the frequency of these floppy modes vanishes as
$\omega\sim\sqrt{\kNNN}$, and the corresponding phonon Green's
function diverges, leading to divergent fluctuation corrections
to the coefficients of $\st0$ in Eq.~\eqref{eq:Fexpansion} as
detailed in Supplementary Information Sec.~III. Keeping leading order terms as
$\kr \rightarrow 0$, we can identify the two phases through the
equations of state,
\begin{align}
	\frac{\partial f(\mathbf{\st0})}{\partial s}
	&\simeq2 ka^2\left(\frac{\tT}{\sqrt{\kr}} +s 	\right)=0 \label{EQ:EOSxx}\\
	\frac{\partial  f(\mathbf{\st0})}{\partial t}
	&\simeq ka^2t
	\left(\kr  + \frac{\lambda\tT}{\sqrt{\kr}}  +\frac{1}{12} \lambda t^{2} \right)=0 ,
	 \label{EQ:EOSxy}
\end{align}
where
\begin{equation}
\tT = \frac{\pi T }{8 k a^2}
\end{equation}
is a unitless reduced temperature. Eq.~(\ref{EQ:EOSxy}) has
three solution for $t$: $t = 0$ corresponding to the square
phase, and two solutions for $t \ne 0$ corresponding to the two
orientations of the rhombic phase. There is only a single
solution for $s$, with $s < 0$, from which we conclude that
both phases exhibit \emph{negative thermal expansion}. The
elastic rigidity and thus the stability of the two phases is
determined by the second derivatives of $F$ with respect to $s$
and $t$. In particular, the reduced shear modulus ($G/k$
where $G$ is the shear modulus) is
\begin{align}\label{EQ:Rordered}
	r =\frac{1}{k} \frac{\partial^2 f(\mathbf{\st0})}{\partial t^2}
	\simeq \kr + \frac{\tT}{\sqrt{\kr}} + \frac{1}{4} \lambda t^2.
\end{align}
To obtain these leading order equations we (i) assume low
$T$, so
only terms singular in $\tau$ as $\tau\to 0$, such as
$t/\sqrt{\tau}$, in the integral of
$\frac{T}{2}\ln\textrm{Det}\tilde{\mathbf{D}}(\mathbf{\dg})$,
and (ii) assume that
\begin{align}\label{EQ:Assu}
	 |s| \sim t^2 \ll \kr \ll 1 ,
\end{align}
the validity of which will be verified below.

As observed in the simulation (Fig.~\ref{FIG:PD}), thermal
fluctuations at $T >0$ stabilize the square relative to the
rhombic phase even for $\kNNN<0$. To understand this phenomenon
within the analytic approach, we use a self-consistent-field
approximation in which $\kr$ is replaced by with its
renormalized value $r$ in the phonon Green's function
$\mathbf{G}_0$  and thus in the denominators on the right hand
sides of Eqs.~(\ref{EQ:EOSxx}), (\ref{EQ:EOSxy}), and
(\ref{EQ:Rordered}). In this approximation, the shear rigidity
of the square ($t=0$) and rhombic ($t\neq 0$) satisfy
\begin{align}\label{EQ:rsquare}
	r =
    \begin{cases}
    \begin{array}{ll}
    \kr + \frac{\tT\lambda}{\sqrt{r}} &\qquad\text{square} \\
    -2 \kr - 2 \frac{\tT\lambda}{\sqrt{r}} & \qquad \text{rhombic}
    \end{array},
    \end{cases}
\end{align}
where we used the equation of state, Eq.~(\ref{EQ:EOSxy}) to
eliminate $t^2$ from Eq.~(\ref{EQ:Rordered}). In the square
phase, Eq.~(\ref{EQ:rsquare}) has a solution $r > 0$, implying
local stability, everywhere except at $\tT = 0,
\kr < 0$ and in the uninteresting limit, $z \rightarrow -
\infty$. This local stability implies that the transition to
the rhombic phase must be first order. In the rhombic phase,
solutions $r>0$ only exist for $\kr < \kr_{c1}$ ($z < z_{c1}$),
where
\begin{equation}
\kr_{c1} = -\frac{3}{2} (\tT\lambda)^{2/3} 
\label{eq:krc1}
\end{equation}

\begin{figure}
\centering
\includegraphics{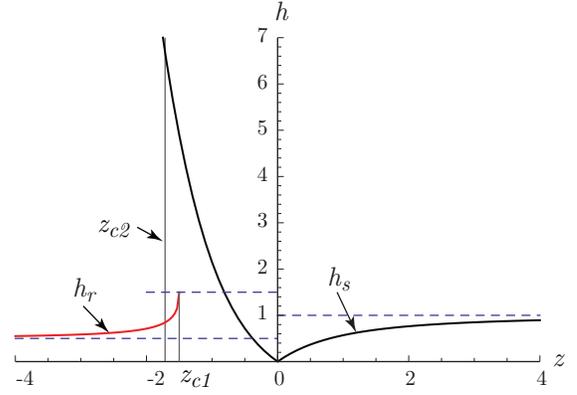}
\caption{(color online) Plots of $h_s(|z|)$ (black) and $h_r(|z|)$ (red)
as a function of $z$.  Note the singular behavior of $h_r(|z|$ in the
vicinity of $z_{c1}$ and the large difference between $h_s$ and $h_r$ at the
first order transition at $z=z_{c2}$.}
\label{fig:scalefn}
\end{figure}

The solutions to Eq.(\ref{EQ:rsquare}) can conveniently be
expressed as scaling functions in the two phases:
\begin{equation}
\frac{|\kr|}{r} = h_{\nu} (|z|), \qquad z = \frac{\kr}{(\tT\lambda)^{2/3}} ,
\label{Eq:scaling1}
\end{equation}
where $\nu=s,\,\,r$  for the square and rhombic phases,
respectively. The scaling functions $h_s(|z|)$ and $h_r(|z|)$
depicted in Fig.~\ref{fig:scalefn} have the following limits
\begin{eqnarray}
h_s(|z|) &\sim  &
\begin{cases}
1 & \qquad z \rightarrow +\infty \\
|z| & \qquad z \rightarrow 0 \\
|z|^3 & \qquad z \rightarrow - \infty
\end{cases} \\ \label{EQ:hsscaling}
h_r(|z|) & \sim &
\begin{cases}
(3/2)-\sqrt{6|z-z_{c1}|} & \qquad z\rightarrow z_{c1}^- \\
1/2 & \qquad z\rightarrow z \rightarrow - \infty,
\end{cases}\label{EQ:hrscaling}
\end{eqnarray}
where $z_{c1}=-3/2$.
The $z^3$ regimes of $h_s$ is in the metastable regime where
the rhombic phase is stable. These results yield the scaling
phase diagram of Fig.~\ref{FIG:PD}.

The phase boundary of this discontinuous transition occurs
along the coexistence line (i.e., equal free-energy line) of
the two phases.  Following Brazovskii, we have already
calculated the limit of metastability of the rhombic phase,
i.e., for the value of $\kNNN=\kr_{c1}$ [Eq.~(\ref{eq:krc1})]
at the local free energy minimum where that phase first
appears. We then calculate the free energy difference between
the two phases, which is evaluated through the following
integral for a given $\kr$,
\begin{align}\label{EQ:DeltaF}
	\Delta F = \int_{0}^{t_{\textrm{rhombic}}}
\frac{d F(\mathbf{\st0})}{d t} dt =\int_{r_0}^{r_1}
\frac{d F(\mathbf{\st0})}{d t} \frac{dt}{dr} dr ,
\end{align}
where we have substituted $r(t)$ for the integral measure. Here
$r_0$ and $r_1$ are the values of $r$ at the minima of $F$
corresponding to the square and the rhombic phases determined
by Eq.~(\ref{EQ:rsquare}). Along the path of this integral,
Eq.~\eqref{EQ:Rordered} is valid, but the equations of
state~(\ref{EQ:EOSxx}) and (\ref{EQ:EOSxy}) and the equation
[Eq.~(\ref{EQ:rsquare})] for $r$ in the rhombic phase are not
satisfied, because they only apply to equilibrium states. The
phase boundary corresponds to the curve of $\kr=\kr_{c2}$ along
which $\Delta F\vert_{\kr = \kr_{c2}}=0$.
As shown in Supplementary Information Sec. IV,
an asymptotic solution valid at low $\kr$, can be obtained
by expanding the equation around $\kr=\kr_{c1}$, assuming that
$\kr_1$ and $\kr_2$ are of the same order of magnitude
(verified below). This yields
\begin{align}\label{EQ:PD}
	\kr_{c2} =- \left(\frac{3}{2}+c \right) (\tT\lambda)^{2/3} = - 1.716 \tT^{2/3} ,
\end{align}
for $\tT\ll 1$ where $c\simeq 0.216$ is a constant.  This
transition line is shown in Fig.~\ref{FIG:PD}. Excellent
agreement between theory and simulation is obtained without any
fitting parameter.

Along the phase boundary, $r_0,r_1\sim \tT^{2/3}>0$, so that
both phases are locally stable. The order parameter for the
transition, $t$, jumps from $0$ to
\begin{align}
	t_{c2} = 3.4 \tT^{1/3}\lambda^{-1/6}
\end{align}
at the transition.  As $T\to 0 $ this discontinuity vanishes,
consistent with the continuous nature of the transition at $T =
0$.  A good agreement between $t$ values in theory and in
simulation is shown in Fig.~\ref{fig:OPNTE}(a).

From Eq.~(\ref{EQ:EOSxx}), the negative thermal expansion
coefficient in both the square and rhombic phases,
\begin{equation}
s= -\frac{\tT}{\sqrt{r}} ,
\end{equation}
is determined by the equation of state, Eq.~\eqref{EQ:EOSxx}.
Equation (\ref{EQ:rsquare}) for $r$ then implies the following
behavior for $s$ in different regions of the phase diagram:  In
the critical region $0<|\kr|<\tT^{2/3}$ of the square phase,
\begin{equation}
s \simeq - \tT^{2/3} \lambda^{-1/3}.
\end{equation}
Deep in the square and rhombic phases, where $0<\tT^{2/3} \ll
|\kr|$,
\begin{equation}
s =
    \begin{cases}
    \begin{array}{ll}
    - \tT/\sqrt{|\tau|} & \qquad \text{square} \\
    -\tT/\sqrt{2|\tau|} & \qquad \text{rhombic}
    \end{array}
    \end{cases} .
\end{equation}
Finally along the coexistence curve in both phases,
$s\sim-\tT/\sqrt{|\kr|} \sim \tT^{2/3}$. These results agree
well with simulation measurement of negative thermal expansion,
as shown in Fig.~\ref{fig:OPNTE}b. In this lattice the negative
thermal expansion behavior results from strong transverse
fluctuations associated with soft modes.

These solutions for $s$ and $t$ verifies that our assumptions
in Eq.~(\ref{EQ:Assu}) are satisfied, provided that $\gr \gg
1$.

\section{Review}

We have presented an analysis of a model based on the
square lattice with NN harmonic and NNN anharmonic springs that
can be tuned at zero temperature from a stable square lattice
through the mechanically unstable NN square lattice to a
highly degenerate zigzag state by changing the coefficient
$\kNNN$ of the harmonic term in the NNN spring from positive
through zero to negative. Using analytic theory, including a
generalization of the Brazovskii theory for the
liquid-to-crystal transition, we investigated the phase diagram
and mechanical properties of this model at $T>0$. The
degeneracy of the zero-$T$ zigzag state is broken by an
order-by-disorder effect, thermal fluctuations drive the
square-to-rhombic phase transition first order,  and the
elastic modulus of the square phase a crossover from being
proportional to $\kNNN$ for $\kNNN\gg k(Tg/k^2)^{2/3} >0$ to
$T^{2/3}$ for $|\kNNN| \ll k(Tg/k^2)^{2/3}$ to $T^2$ for
$\kappa \ll - k(Tg/k^2)^{2/3}<0$ as function of the scaling
variable $z=\tau/(\tilde{T} \lambda)^{2/3} = \sim
(\kNNN/k)/(Tg/k^2)^{2/3}$. This behavior arises because the
spectrum of the NN square lattice with $N$ sites exhibits
$\sqrt{N}$ zero modes on a one-dimensional manifold in the
Brillouin Zone. Other lattices such as the 2D kagome lattice,
the 3D simple cubic lattice, and the 3D pyrochlore and
$\beta$-cristobalite \cite{Hammonds1996} lattices have similar
spectra, and it is our expectation that generalizations of our
model to these lattices will exhibit similar behavior.  It is
also likely that our model can inform us about more physically
realistic models in which interactions lead to spectra with a
large set of modes with small but not zero frequency.

\textit{Acknowledgments -- } A.S. gatefully acknowledges discussions with P. A. Rikvold, Gregory Brown, Shengnan Huang and Andrea J. Liu. This work was supported in part by the NSF under grants DMR-1104707 and DMR-1120901 (TCL), DMR-1207026 (AS), grant DGAPA IN-110613 (CIM) and by the Georgia Institute of Technology (AS).

\bibliographystyle{naturemag}

\begin{thebibliography}{10}
\expandafter\ifx\csname url\endcsname\relax
  \def\url#1{\texttt{#1}}\fi
\expandafter\ifx\csname urlprefix\endcsname\relax\def\urlprefix{URL }\fi
\providecommand{\bibinfo}[2]{#2}
\providecommand{\eprint}[2][]{\url{#2}}

\bibitem{FolkSch1976}
\bibinfo{author}{Folk, R.}, \bibinfo{author}{Iro, H.} \&
  \bibinfo{author}{Schwabl, F.}
\newblock \bibinfo{title}{Critical elastic phase transtions}.
\newblock \emph{\bibinfo{journal}{Z. Physik B}} \textbf{\bibinfo{volume}{25}},
  \bibinfo{pages}{69--81} (\bibinfo{year}{1976}).

\bibitem{Cowley1980}
\bibinfo{author}{Cowley, R.~A.}
\newblock \bibinfo{title}{Structural phase-transitions i: Landau theory}.
\newblock \emph{\bibinfo{journal}{Advances in Physics}}
  \textbf{\bibinfo{volume}{29}}, \bibinfo{pages}{1--110}
  (\bibinfo{year}{1980}).

\bibitem{Bruce1980}
\bibinfo{author}{Bruce, A.}
\newblock \bibinfo{title}{Structural phase transitions ii: Static critical
  behaviour}.
\newblock \emph{\bibinfo{journal}{Advances in Physics}}
  \textbf{\bibinfo{volume}{29}}, \bibinfo{pages}{111--217}
  (\bibinfo{year}{1980}).

\bibitem{Fujimoto2005}
\bibinfo{author}{Fujimoto, M.}
\newblock \emph{\bibinfo{title}{The Physics of Structural Phase Transitions}}
  (\bibinfo{publisher}{Springer}, \bibinfo{year}{2005}), \bibinfo{edition}{2nd}
  edn.

\bibitem{Maxwell1864}
\bibinfo{author}{Maxwell, J.~C.}
\newblock \bibinfo{title}{On the calculation of the equilibrium and stiffness
  of frames}.
\newblock \emph{\bibinfo{journal}{Philos. Mag.}} \textbf{\bibinfo{volume}{27}},
  \bibinfo{pages}{294} (\bibinfo{year}{1864}).

\bibitem{Note1}
\bibinfo{note}{The term \protect \emph {isostatic} is often incorrectly used to
  describe any system with $z=z_c$. Finite Isostatic lattices have $z=z_c^N=z_c
  -d(d+1)/N$ and no states of self stress - See~\cite {Calladine1978} for
  example. There is no universally accepted definition of \protect \emph
  {isostatic} in lattices with periodic boundary conditions, but one such as
  the square lattice with many states of self-stress is surely not isostatic.}

\bibitem{Heyman1999}
\bibinfo{author}{Heyman, J.}
\newblock \emph{\bibinfo{title}{The Science of Structural Engineering}}
  (\bibinfo{publisher}{Cengage Learning}, \bibinfo{address}{Stamford CT},
  \bibinfo{year}{2005}).

\bibitem{Kassimali2005}
\bibinfo{author}{Kassimali, A.}
\newblock \emph{\bibinfo{title}{Structural Analysis}}
  (\bibinfo{publisher}{Cengage Learning}, \bibinfo{address}{Stamford CT},
  \bibinfo{year}{2005}).

\bibitem{Feng1984}
\bibinfo{author}{Feng, S.} \& \bibinfo{author}{Sen, P.~N.}
\newblock \bibinfo{title}{Percolation on elastic networks: New exponent and
  threshold}.
\newblock \emph{\bibinfo{journal}{Phys. Rev. Lett.}}
  \textbf{\bibinfo{volume}{52}}, \bibinfo{pages}{216--219}
  (\bibinfo{year}{1984}).

\bibitem{FengLob1984}
\bibinfo{author}{Feng, S.}, \bibinfo{author}{Sen, P.~N.},
  \bibinfo{author}{Halperin, B.~I.} \& \bibinfo{author}{Lobb, C.~J.}
\newblock \bibinfo{title}{Percolation on two-dimensional elastic networks with
  rotationally invariant bond-bending forces}.
\newblock \emph{\bibinfo{journal}{Phys. Rev. B}} \textbf{\bibinfo{volume}{30}},
  \bibinfo{pages}{5386--5389} (\bibinfo{year}{1984}).

\bibitem{Jacobs1995}
\bibinfo{author}{Jacobs, D.~J.} \& \bibinfo{author}{Thorpe, M.~F.}
\newblock \bibinfo{title}{Generic rigidity percolation: The pebble game}.
\newblock \emph{\bibinfo{journal}{Phys. Rev. Lett.}}
  \textbf{\bibinfo{volume}{75}}, \bibinfo{pages}{4051--4054}
  (\bibinfo{year}{1995}).

\bibitem{Liu1998}
\bibinfo{author}{Liu, A.~J.} \& \bibinfo{author}{Nagel, S.~R.}
\newblock \bibinfo{title}{Jamming is not just cool any more}.
\newblock \emph{\bibinfo{journal}{Nature}} \textbf{\bibinfo{volume}{396}},
  \bibinfo{pages}{21} (\bibinfo{year}{1998}).

\bibitem{Wyart2005a}
\bibinfo{author}{Wyart, M.}
\newblock \bibinfo{title}{{On the rigidity of amorphous solids}}.
\newblock \emph{\bibinfo{journal}{Ann. Phys. Fr}}
  \textbf{\bibinfo{volume}{30}}, \bibinfo{pages}{1--96} (\bibinfo{year}{2005}).

\bibitem{LiuNag2010a}
\bibinfo{author}{Liu, A.~J.} \& \bibinfo{author}{Nagel, S.~R.}
\newblock \emph{\bibinfo{title}{The Jamming Transition and the Marginally
  Jammed Solid}}, vol.~\bibinfo{volume}{1} of \emph{\bibinfo{series}{Annual
  Review of Condensed Matter Physics}}, \bibinfo{pages}{347--369}
  (\bibinfo{year}{2010}).

\bibitem{Elson1988}
\bibinfo{author}{Elson, E.~L.}
\newblock \bibinfo{title}{Cellular mechanism as an indicator of csk structure
  and function}.
\newblock \emph{\bibinfo{journal}{Annu. Rev. Biophys. Biophys. Chem}}
  \textbf{\bibinfo{volume}{17}}, \bibinfo{pages}{397--430}
  (\bibinfo{year}{1988}).

\bibitem{Kasza2007}
\bibinfo{author}{Kasza, K.} \emph{et~al.}
\newblock \bibinfo{title}{The cell as a material}.
\newblock \emph{\bibinfo{journal}{Curr. Opin. Cell Biol.}}
  \textbf{\bibinfo{volume}{19}}, \bibinfo{pages}{101--107}
  (\bibinfo{year}{2007}).

\bibitem{Alberts2008}
\bibinfo{author}{Alberts, B.} \emph{et~al.}
\newblock \emph{\bibinfo{title}{Molecular Biology of the Cell}}
  (\bibinfo{publisher}{Garland, New York}, \bibinfo{year}{2008}),
  \bibinfo{edition}{4th} edn.

\bibitem{Janmey1990}
\bibinfo{author}{Janmey, P.} \emph{et~al.}
\newblock \bibinfo{title}{Resemblance of actin-binding actin gels to
  covanlently cross-linked networks}.
\newblock \emph{\bibinfo{journal}{Nature}} \textbf{\bibinfo{volume}{345}}
  (\bibinfo{year}{1990}).

\bibitem{BroederszMac2014}
\bibinfo{author}{Broedersz, C.~P.} \& \bibinfo{author}{MacKintosh, F.~C.}
\newblock \bibinfo{title}{Modeling semiflexible polymer networks}.
\newblock \emph{\bibinfo{journal}{arXiv:1404.4332}}  (\bibinfo{year}{2014}).

\bibitem{Phillips1981}
\bibinfo{author}{Phillips, J.~C.}
\newblock \bibinfo{title}{Topology of covalent non-crystalline solids .2.
  medium-range order in chalcogenide alloys and a-si(ge)}.
\newblock \emph{\bibinfo{journal}{J. Non-Cryst. Solids}}
  \textbf{\bibinfo{volume}{43}}, \bibinfo{pages}{37--77}
  (\bibinfo{year}{1981}).

\bibitem{Thorpe1983}
\bibinfo{author}{Thorpe, M.}
\newblock \bibinfo{title}{Continuous deformations in random networks}.
\newblock \emph{\bibinfo{journal}{Journal of Non-Crystalline Solids}}
  \textbf{\bibinfo{volume}{57}}, \bibinfo{pages}{355 -- 370}
  (\bibinfo{year}{1983}).

\bibitem{Souslov2009}
\bibinfo{author}{Souslov, A.}, \bibinfo{author}{Liu, A.~J.} \&
  \bibinfo{author}{Lubensky, T.~C.}
\newblock \bibinfo{title}{Elasticity and response in nearly isostatic periodic
  lattices}.
\newblock \emph{\bibinfo{journal}{Physical Review Letters}}
  \textbf{\bibinfo{volume}{103}}, \bibinfo{pages}{205503}
  (\bibinfo{year}{2009}).

\bibitem{Pieranski1983}
\bibinfo{author}{Pieranski, P.}, \bibinfo{author}{Strzelecki, L.} \&
  \bibinfo{author}{Pansu, B.}
\newblock \bibinfo{title}{Thin colloidal crystals}.
\newblock \emph{\bibinfo{journal}{Physical Review Letters}}
  \textbf{\bibinfo{volume}{50}}, \bibinfo{pages}{900--903}
  (\bibinfo{year}{1983}).

\bibitem{Han2008}
\bibinfo{author}{Han, Y.} \emph{et~al.}
\newblock \bibinfo{title}{Geometric frustration in buckled colloidal
  monolayers}.
\newblock \emph{\bibinfo{journal}{Nature}} \textbf{\bibinfo{volume}{456}},
  \bibinfo{pages}{898--903} (\bibinfo{year}{2008}).

\bibitem{Shokef2011}
\bibinfo{author}{Shokef, Y.}, \bibinfo{author}{Souslov, A.} \&
  \bibinfo{author}{Lubensky, T.~C.}
\newblock \bibinfo{title}{Order by disorder in the antiferromagnetic ising
  model on an elastic triangular lattice}.
\newblock \emph{\bibinfo{journal}{Proceedings of the National Academy of
  Sciences}} \textbf{\bibinfo{volume}{108}}, \bibinfo{pages}{11804--11809}
  (\bibinfo{year}{2011}).

\bibitem{DennisonMac2013}
\bibinfo{author}{Dennison, M.}, \bibinfo{author}{Sheinman, M.},
  \bibinfo{author}{Storm, C.} \& \bibinfo{author}{MacKintosh, F.~C.}
\newblock \bibinfo{title}{Fluctuation-stabilized marginal networks and
  anomalous entropic elasticity}.
\newblock \emph{\bibinfo{journal}{Physical Review Letters}}
  \textbf{\bibinfo{volume}{111}}, \bibinfo{pages}{095503}
  (\bibinfo{year}{2013}).

\bibitem{IkedaBir2013}
\bibinfo{author}{Ikeda, A.}, \bibinfo{author}{Berthier, L.} \&
  \bibinfo{author}{Biroli, G.}
\newblock \bibinfo{title}{Dynamic criticality at the jamming transition}.
\newblock \emph{\bibinfo{journal}{Journal of Chemical Physics}}
  \textbf{\bibinfo{volume}{138}}, \bibinfo{pages}{12a507}
  (\bibinfo{year}{2013}).

\bibitem{Villain1980}
\bibinfo{author}{{Villain, J.}}, \bibinfo{author}{{Bidaux, R.}},
  \bibinfo{author}{{Carton, J.-P.}} \& \bibinfo{author}{{Conte, R.}}
\newblock \bibinfo{title}{Order as an effect of disorder}.
\newblock \emph{\bibinfo{journal}{J. Phys. France}}
  \textbf{\bibinfo{volume}{41}}, \bibinfo{pages}{1263--1272}
  (\bibinfo{year}{1980}).

\bibitem{Shender1982}
\bibinfo{author}{Shender, E.}
\newblock \bibinfo{title}{Anitferromagnetic garnets with fluctuationally
  interacting sublattices}.
\newblock \emph{\bibinfo{journal}{Sov. Phys. JETP}}
  \textbf{\bibinfo{volume}{56}}, \bibinfo{pages}{178--184}
  (\bibinfo{year}{1982}).

\bibitem{Henley1987}
\bibinfo{author}{Henley, C.~L.}
\newblock \bibinfo{title}{Ordering by disorder: Ground state selection in fcc
  vector antiferromagnets}.
\newblock \emph{\bibinfo{journal}{Journal of Applied Physics}}
  \textbf{\bibinfo{volume}{61}}, \bibinfo{pages}{3962--3964}
  (\bibinfo{year}{1987}).

\bibitem{Henley1989}
\bibinfo{author}{Henley, C.~L.}
\newblock \bibinfo{title}{Ordering due to disorder in a frustrated vector
  antiferromagnet}.
\newblock \emph{\bibinfo{journal}{Phys. Rev. Lett.}}
  \textbf{\bibinfo{volume}{62}}, \bibinfo{pages}{2056--2059}
  (\bibinfo{year}{1989}).

\bibitem{Chubukov1992}
\bibinfo{author}{Chubukov, A.}
\newblock \bibinfo{title}{Order from disorder in a kagomÃ© antiferromagnet}.
\newblock \emph{\bibinfo{journal}{Phys. Rev. Lett.}}
  \textbf{\bibinfo{volume}{69}}, \bibinfo{pages}{832--835}
  (\bibinfo{year}{1992}).

\bibitem{Reimers1993}
\bibinfo{author}{Reimers, J.~N.} \& \bibinfo{author}{Berlinsky, A.~J.}
\newblock \bibinfo{title}{Order by disorder in the classical heisenberg kagomÃ©
  antiferromagnet}.
\newblock \emph{\bibinfo{journal}{Phys. Rev. B}} \textbf{\bibinfo{volume}{48}},
  \bibinfo{pages}{9539--9554} (\bibinfo{year}{1993}).

\bibitem{Bergman2007}
\bibinfo{author}{{Bergman, Doron}}, \bibinfo{author}{{Alicea, Jason}},
  \bibinfo{author}{{Gull, Emanuel}}, \bibinfo{author}{{Trebst, Simon}} \&
  \bibinfo{author}{{Balents, Leon}}.
\newblock \bibinfo{title}{Order-by-disorder and spiral spin-liquid in
  frustrated diamond-lattice antiferromagnets}.
\newblock \emph{\bibinfo{journal}{Nat Phys}} \textbf{\bibinfo{volume}{3}},
  \bibinfo{pages}{487--491} (\bibinfo{year}{2007}).

\bibitem{RubinsteinBas1992}
\bibinfo{author}{Rubinstein, M.}, \bibinfo{author}{Leibler, L.} \&
  \bibinfo{author}{Bastide, J.}
\newblock \bibinfo{title}{Giant fluctuations of cross-linked positions in
  gels}.
\newblock \emph{\bibinfo{journal}{Physical Review Letters}}
  \textbf{\bibinfo{volume}{68}}, \bibinfo{pages}{405--407}
  (\bibinfo{year}{1992}).

\bibitem{Barriere1995}
\bibinfo{author}{Barriere, B.}
\newblock \bibinfo{title}{Elatic moduli of 2d grafted tethered membranes}.
\newblock \emph{\bibinfo{journal}{Journal De Physique I}}
  \textbf{\bibinfo{volume}{5}}, \bibinfo{pages}{389--398}
  (\bibinfo{year}{1995}).

\bibitem{Plischke1998}
\bibinfo{author}{Plischke, M.} \& \bibinfo{author}{Jo{\'o}s, B.}
\newblock \bibinfo{title}{Entropic elasticity of diluted central force
  networks}.
\newblock \emph{\bibinfo{journal}{Physical review letters}}
  \textbf{\bibinfo{volume}{80}}, \bibinfo{pages}{4907} (\bibinfo{year}{1998}).

\bibitem{TessierDis2003}
\bibinfo{author}{Tessier, F.}, \bibinfo{author}{Boal, D.~H.} \&
  \bibinfo{author}{Discher, D.~E.}
\newblock \bibinfo{title}{Networks with fourfold connectivity in two
  dimensions}.
\newblock \emph{\bibinfo{journal}{Physical Review E}}
  \textbf{\bibinfo{volume}{67}}, \bibinfo{pages}{011903}
  (\bibinfo{year}{2003}).

\bibitem{Brazovskii1975}
\bibinfo{author}{Brazovskii, S.~A.}
\newblock \bibinfo{title}{Phase-transition of an isotropic system to an
  inhomogeneous state}.
\newblock \emph{\bibinfo{journal}{Zh Eksp Teor Fiz}}
  \textbf{\bibinfo{volume}{68}}, \bibinfo{pages}{175--185}
  (\bibinfo{year}{1975}).

\bibitem{SwiftHoh1977}
\bibinfo{author}{Swift, J.} \& \bibinfo{author}{Hohenberg, P.~C.}
\newblock \bibinfo{title}{Hydrodynamic fluctuations at convective instability}.
\newblock \emph{\bibinfo{journal}{Physical Review A}}
  \textbf{\bibinfo{volume}{15}}, \bibinfo{pages}{319--328}
  (\bibinfo{year}{1977}).

\bibitem{ChenLub1976}
\bibinfo{author}{Chen, J.~H.} \& \bibinfo{author}{Lubensky, T.~C.}
\newblock \bibinfo{title}{Landau-ginzburg mean-field theory for nematic to
  smectic-c and nematic to smectic-a phase- transitions}.
\newblock \emph{\bibinfo{journal}{Physical Review A}}
  \textbf{\bibinfo{volume}{14}}, \bibinfo{pages}{1202--1207}
  (\bibinfo{year}{1976}).

\bibitem{Swift1976}
\bibinfo{author}{Swift, J.}
\newblock \bibinfo{title}{Fluctuations near nematic-smectic-c
  phase-transition}.
\newblock \emph{\bibinfo{journal}{Physical Review A}}
  \textbf{\bibinfo{volume}{14}}, \bibinfo{pages}{2274--2277}
  (\bibinfo{year}{1976}).

\bibitem{Calladine1978}
\bibinfo{author}{Calladine, C.~R.}
\newblock \bibinfo{title}{Buckminster fuller "tensegrity" structures and clerk
  maxwell ruels for the construction of stiff frames}.
\newblock \emph{\bibinfo{journal}{International Journal of Solids and
  Structures}} \textbf{\bibinfo{volume}{14}}, \bibinfo{pages}{161--172}
  (\bibinfo{year}{1978}).

\bibitem{Sun2012}
\bibinfo{author}{Sun, K.}, \bibinfo{author}{Souslov, A.}, \bibinfo{author}{Mao,
  X.} \& \bibinfo{author}{Lubensky, T.~C.}
\newblock \bibinfo{title}{Surface phonons, elastic response, and conformal
  invariance in twisted kagome lattices}.
\newblock \emph{\bibinfo{journal}{Proceedings of the National Academy of
  Sciences}} \textbf{\bibinfo{volume}{109}}, \bibinfo{pages}{12369--12374}
  (\bibinfo{year}{2012}).

\bibitem{Kane2014}
\bibinfo{author}{Kane, C.} \& \bibinfo{author}{Lubensky, T.}
\newblock \bibinfo{title}{Topological boundary modes in isostatic lattices}.
\newblock \emph{\bibinfo{journal}{Nature Physics}}
  \textbf{\bibinfo{volume}{10}}, \bibinfo{pages}{39--45}
  (\bibinfo{year}{2014}).

\bibitem{Schmidt1996}
\bibinfo{author}{Schmidt, M.} \& \bibinfo{author}{L{\"o}wen, H.}
\newblock \bibinfo{title}{Freezing between two and three dimensions}.
\newblock \emph{\bibinfo{journal}{Physical Review Letters}}
  \textbf{\bibinfo{volume}{76}}, \bibinfo{pages}{4552--4555}
  (\bibinfo{year}{1996}).

\bibitem{Schmidt1997}
\bibinfo{author}{Schmidt, M.} \& \bibinfo{author}{L{\"o}wen, H.}
\newblock \bibinfo{title}{Phase diagram of hard spheres confined between two
  parallel plates}.
\newblock \emph{\bibinfo{journal}{Physical Review E}}
  \textbf{\bibinfo{volume}{55}}, \bibinfo{pages}{7228--7241}
  (\bibinfo{year}{1997}).

\bibitem{Shokef2009}
\bibinfo{author}{Shokef, Y.} \& \bibinfo{author}{Lubensky, T.~C.}
\newblock \bibinfo{title}{Stripes, zigzags, and slow dynamics in buckled hard
  spheres}.
\newblock \emph{\bibinfo{journal}{Phys. Rev. Lett.}}
  \textbf{\bibinfo{volume}{102}}, \bibinfo{pages}{048303}
  (\bibinfo{year}{2009}).

\bibitem{Frenkel2001}
\bibinfo{author}{Frenkel, D.} \& \bibinfo{author}{Smit, B.}
\newblock \emph{\bibinfo{title}{Understanding Molecular Simulations}}
  (\bibinfo{publisher}{Academic Press, San Diego}, \bibinfo{year}{2001}).

\bibitem{Binder1981a}
\bibinfo{author}{Binder, K.}
\newblock \bibinfo{title}{Critical properties from monte carlo coarse graining
  and renormalization}.
\newblock \emph{\bibinfo{journal}{Physical Review Letters}}
  \textbf{\bibinfo{volume}{47}}, \bibinfo{pages}{693--696}
  (\bibinfo{year}{1981}).

\bibitem{Mao2010}
\bibinfo{author}{Mao, X.}, \bibinfo{author}{Xu, N.} \&
  \bibinfo{author}{Lubensky, T.~C.}
\newblock \bibinfo{title}{Soft modes and elasticity of nearly isostatic
  lattices: Randomness and dissipation}.
\newblock \emph{\bibinfo{journal}{Phys. Rev. Lett.}}
  \textbf{\bibinfo{volume}{104}}, \bibinfo{pages}{085504}
  (\bibinfo{year}{2010}).

\bibitem{Note2}
\bibinfo{note}{This form is similar to Eq.~\ref {eq:sp}, except that here we
  expand around uniformly deformed lattice of continuously varying $\Lambda $,
  rather than about a zigzagging state.}

\bibitem{Hammonds1996}
\bibinfo{author}{Hammonds, K.~D.}, \bibinfo{author}{Dove, M.~T.},
  \bibinfo{author}{Giddy, A.~P.}, \bibinfo{author}{Heine, V.} \&
  \bibinfo{author}{Winkler, B.}
\newblock \bibinfo{title}{Rigid-unit phonon modes and structural phase
  transitions in framework silicates}.
\newblock \emph{\bibinfo{journal}{American Mineralogist}}
  \textbf{\bibinfo{volume}{81}}, \bibinfo{pages}{1057--1079}
  (\bibinfo{year}{1996}).

\end{thebibliography}

\newpage
\appendix

\begin{widetext}
\section{Plaquette Ground State}
The minimum of $V_{\mathrm{PL}}$ in Eq.~(4.5) in the main text can be 
found analytically by assuming a rhombic plaquette and solving the set of
equations given by $\partial V_{\mathrm{PL}}/ \partial d_1 = 0$
and  $\partial V_{\mathrm{PL}}/ \partial d_2 = 0$ for $\tau$ and $\lambda$.
These equations are linear, and may be written in terms of the 
plaquette side $b$ and the inner angle $\alpha$, where $d_{1,2} = \sqrt{2} b \sqrt{1 \pm \sin\alpha}$.
The solutions have the form
\begin{align}
\tau(b, \alpha) & = \frac{(b -1) \left\lbrack b^2 \left(3 \cos\alpha +b \cos\frac{3 \alpha }{2} -3 b \cos\frac{\alpha }{2}\right) 
- 1\right)}{4 b \left(b \cos\frac{\alpha }{2} - 1\right) \left(1-2 b \cos\frac{\alpha }{2}+b^2 \cos\alpha\right)}, \nonumber \\
\lambda(b, \alpha) & = \frac{3 (b-1)}{4 b\left(b \cos\frac{\alpha}{2}- 1\right) \left(1-2 b \cos\frac{\alpha }{2}+b^2 \cos\alpha\right)}.
\end{align}
It was verified numerically that the rhombic plaquette with side length $b$ minimizes the potential energy
for the range of parameters considered in this work relative to plaquettes with sides of unequal length.

\section{Expansion of lattice Hamiltonian at deformed reference states}
For a generic lattice with pair-wise potentials, we can write the Hamiltonian as 
\begin{align}
	H=\sum_{b} V_b \left( \vert \vec{R}_b\vert - \vert \vec{R}_{0b}\vert \right)
\end{align}
where $b$ labels bonds $\{ i,j\}$, $V_b$ is the interaction potential of the bond, and 
\begin{align}
	\vec{R}_{0b} &= \vec{R}_{0i} - \vec{R}_{0j}\nonumber\\
	\vec{R}_b &= \vec{R}_i - \vec{R}_j 
\end{align}
are the bond vectors in the reference and target states.

We consider a macroscopic deformation $\Lambda$ (corresponding to $\stm$ in the continuum theory) and fluctuations $\vec{u}'$, so the deformation of an arbitrary site can be written as
\begin{align}
	\vec{R}_{0i} \to \vec{R}_i = \mathbf{\Lambda}\cdot \vec{R}_{0i} + \vec{u}'_{i}.
\end{align}
Thus for a bond,
\begin{align}
	\vec{R}_{0b} \to \vec{R}_b = \mathbf{\Lambda}\cdot \vec{R}_{0b} + \vec{u}'_{b}.
\end{align}

The change of bond length can then be expanded for small $\vec{u}'_b$
\begin{align}
	\vert \vec{R}_b\vert - \vert \vec{R}_{0b}\vert
	= \vert \mathbf{\Lambda}\cdot \vec{R}_{0b} \vert - \vert \vec{R}_{0b}\vert
	+ \tilde{u}'_{b\parallel} +\frac{(\tilde{u}'_{b\perp})^2}{2\vert \mathbf{\Lambda}\cdot \vec{R}_{0b} \vert}
	+\cdots
\end{align}
where
\begin{align}
	\tilde{u}'_{b\parallel}&= \vec{u}'_{b} \cdot \hat{t}_b \nonumber\\
	(\tilde{u}'_{b\perp})^2 &= \vec{u}'_{b} \cdot (\mathbf{I} - \hat{t}_b\hat{t}_b) \cdot  \vec{u}'_{b}
\end{align}
with
\begin{align}
	\hat{t}_b =\frac{\mathbf{\Lambda}\cdot \vec{R}_{0b}}{\vert \mathbf{\Lambda}\cdot \vec{R}_{0b}\vert} .
\end{align}
The terms we kept are exact to $O(\tilde{u}^2)$.

The expansion of the potential of a bond is then
\begin{align}
	V_b = V_{b\Lambda} +
	\left( \tilde{u}'_{b\parallel} +\frac{(\tilde{u}'_{b\perp})^2}{2\vert \mathbf{\Lambda}\cdot \vec{R}_{0b} \vert}\right) V'_{b\Lambda} 
	+ 
	\left( \tilde{u}'_{b\parallel} +\frac{(\tilde{u}_{b\perp})^2}{2\vert \mathbf{\Lambda}\cdot \vec{R}_{0b} \vert}\right)^2 \frac{1}{2}V''_{b\Lambda} 
	+ \ldots
\end{align}
where $V_{b\Lambda}$, $V'_{b\Lambda}$, $V''_{b\Lambda}$ are the potential and its derivatives at the macroscopic deformation value
\begin{align}
	\mathbf{\Lambda}\cdot \vec{R}_{0b}.
\end{align}
We can then sort terms in powers of $\tilde{u}'$
\begin{align}
	V_b = V_{b\Lambda} + V'_{b\Lambda}\tilde{u}'_{b\parallel} +
	\left( \frac{V'_{b\Lambda} }{2\vert \mathbf{\Lambda}\cdot \vec{R}_{0b} \vert} (\tilde{u}'_{b\perp})^2
	+\frac{V''_{b\Lambda}}{2}  (\tilde{u}'_{b\parallel})^2\right) + O((\tilde{u}')^3) .
\end{align}

Therefore the total Hamiltonian of the lattice is
\begin{align}
	H(\Lambda,\vec{u}') = H_0(\Lambda) + H^{(2)}(\Lambda,\vec{u}') +O((\tilde{u}')^3) ,
\end{align}
where $H_0(\Lambda)$ is the energy for reference state with the uniform deformation ($\Lambda$), and 
\begin{align}
	H^{(2)} (\Lambda,\vec{u}') = \frac{1}{V } \sum_q \vec{u}'_q \cdot \mathbf{D}_q(\Lambda) \cdot \vec{u}'_{-q}
\end{align}
is the additional potential energy coming from fluctuations around the uniformly deformed reference state.  Here we express it in momentum space, $v_0$ is the area of the unit cell, and
\begin{align}\label{EQ:DLatt}
	\mathbf{D}_q (\Lambda) = \sum_{B} 2\lbrack 1-\cos(q\cdot R_{0B}) \rbrack \mathbf{A}_B
\end{align}
with
\begin{align}
	A_{B} = \frac{V'_{B\Lambda} }{2\vert \mathbf{\Lambda}\cdot \vec{R}_{0B} \vert} \mathbf{I}
	+ \left( \frac{V''_{B\Lambda}}{2} - \frac{V'_{B\Lambda} }{2\vert \mathbf{\Lambda}\cdot \vec{R}_{0B} \vert} \right)
	\mathbf{\Lambda}\cdot 
	\left( \frac{\vec{R}_{0B}\vec{R}_{0B}}{\vert \mathbf{\Lambda}\cdot \vec{R}_{0B} \vert^2} \right)
	\cdot\mathbf{\Lambda}^{T}
\end{align}
and $\sum_{B}$ represent the sum over all bonds in one unit cell.  From this we can write the dynamical matrix $D$ as a sum of two parts
\begin{align}
	\mathbf{D}_q (\Lambda) = v_q(\Lambda) \mathbf{I} + \mathbf{\Lambda}\cdot \mathbf{M}_q(\Lambda) \mathbf{\Lambda}^{T} .
\end{align}
We can then calculate the expression for $\mathbf{D}_q (\Lambda)$ for the special case of the square lattice model with NN harmonic springs of spring constant $k$ and NNN springs with the potential
\begin{align}
	V_{\textrm{NNN}}(\Delta R) = \frac{\kappa}{2} \Delta R ^2 + \frac{g}{4!} \Delta R ^4 .
\end{align}

\section{Lattice Free Energy}
From the Hamiltonian derived in the previous section, one can calculate the free energy of a state with a uniform deformation $\Lambda$ by integrating out small fluctuations $\vec{u}'$ around this state.  This lead to 
\begin{align}
	F(\Lambda)& = - T \ln \int \mathcal{D}\vec{u}' e^{-H(\Lambda,\vec{u}')/ T} \nonumber\\
	&= H_0(\Lambda) +  \frac{1}{2} T \ln \det \left\lbrack \mathbf{D}_q (\Lambda) 
	\right\rbrack .
\end{align}
Using the form of $\mathbf{D}_q (\Lambda)$ and the equality $\det(I+AB)=\det(I+BA)$ we find that $F$ only depends on the rotationally invariant combination of the uniform deformation $\stm \equiv (\Lambda^{T}\Lambda -I)/2$:
\begin{align}
	F(\stm)&
	= H(\stm) +  \frac{1}{2} T \ln \det \left\lbrack v(\mathbf{\stm}) \mathbf{I} + 
		(\mathbf{I}+2\mathbf{\stm}) \mathbf{M}(\stm)
	\right\rbrack ,
\end{align}
where $g$ starts from $O(\stm)$ and $M$ starts from $O(1)$.  This confirms the Ward identity in this problem.

To analyze the transition we then expand $F$ as a series of $\stm$.  Because the dynamical matrix can be expanded as
\begin{align}
	\mathbf{D} = \mathbf{D}_0 +\mathbf{A}(\stm)
\end{align}
where $ \mathbf{D}_0 = \mathbf{D}\vert_{\mathbf{\st0} = 0}$ is the dynamical matrix of the undeformed state.
The free energy is then
\begin{align}\label{eq:Fexpansion}
	F(\mathbf{\stm}) = H_0(\mathbf{\stm}) + \frac{T}{2}\textrm{Tr} \ln
	 \mathbf{D}_0 \left\lbrack \mathbf{I}+\mathbf{G}_0 \cdot \mathbf{A}(\mathbf{\stm})\right\rbrack ,
\end{align}
where $\mathbf{G}_0 \equiv \mathbf{D}_0 ^{-1}$ is the phonon Green's function in the undeformed state. The expansion of $F$ at small $\mathbf{\stm}$ thus follows from this,
\begin{align}\label{EQ:FEExp}
	F(\stm)
	&= H(\stm) +  \frac{1}{2} T \ln \det  \mathbf{D_0}
	+ \frac{1}{2} T \ln \det \big\lbrack \mathbf{I}+ \mathbf{G_0 \cdot A}
	 \big\rbrack \nonumber\\
	&\simeq H(\stm) +  \frac{1}{2} T \ln \det  \mathbf{D_0}
	+ \frac{1}{2} T \, \textrm{Tr}  \big\lbrack  \mathbf{G_0 \cdot A}
	 -\frac{1}{2}\mathbf{G_0 \cdot A \cdot G_0 \cdot A} 	 + \frac{1}{3} \mathbf{G_0 \cdot A \cdot G_0 \cdot A \cdot G_0 \cdot A} 
	\big\rbrack
\end{align}

The above free energy can be calculated by performing integrals in momentum space.  Because we are interested in characterizing the square-rhombic phase transition, where $\tau\equiv\kappa/k$ is small, we can expand our results at small $\tau$.  In this limit, floppy modes of frequency $\sqrt{\tau}$ lie along $q_x$ and $q_y$ axes, and to lowest order in $\tau$, the Green's function can be approximated by
\begin{align}
	\mathbf{D}_0 = \left( \begin{array}{cc}
	k q_x^2 +\kappa q_y^2 & 0 \\
	0 & k q_y^2 + \kappa q_x^2
	\end{array} \right).
\end{align}
Integrals involving this Green's functions can be evaluated following the calculation discussed in Ref.[40] (add the kagome lattice ref here which contain more detail).  We also verified that to leading order in small $\tau$, integrals done using this approximation agree with exact results.

\section{Identify the transition}
The square and the rhombic phases can be written in terms of a uniform deformation
\begin{align}\label{EQ:Lambda}
	\mathbf{\Lambda} = \left( \begin{array}{cc}
	1+s & t \\
	0 & 1+s+w
	\end{array} \right),
\end{align}
where $s$ denote a hydrostatic expansion, $t$ measures simple shear, and $w$ denote an anisotropic expansion in the $y$ direction.  The shear $t$ is the order parameter of the transition: $t=0$ in the square phase and $t\ne 0$ in the rhombic phase.  The free energy can then be written in terms of deformations $\{t,s,w\}$.

At $T=0$, one can solve for $\{t,s,w\}$ by minimizing $H_0$.  For $\kappa>0$ the solution is simply $\{t=0, s=0, w=0\}$ corresponding to the square lattice.  For $\kappa<0$ we find, in terms of the dimensionless variables $\tau\equiv\kappa/k$ and $\lambda\equiv g a^2/k$,
\begin{align}\label{EQ:T0EOS}
	t&=\pm \sqrt{12\vert \tau\vert/\lambda}+O(\tau^{3/2}) ,\nonumber\\
	s&= O(\tau^{2}) ,\nonumber\\
	w&=-6\vert \tau\vert/\lambda 
\end{align}
corresponding to a shear, associated with a small vertical shrink to keep the length of the vertical bonds unchanged.  Because $t\propto \vert \tau\vert^{1/2}$ this is a continuous transition at $T=0$.

At $T>0$ the equilibrium state is determined by the free energy $F$ as given in Eq.~\eqref{EQ:FEExp}.  In particular, we can identify the transition from the equations of state
\begin{align}
	\frac{\partial F(\stm)}{\partial t} =0, \quad \frac{\partial F(\stm)}{\partial s} =0, \quad \frac{\partial F(\stm)}{\partial w} =0 .
\end{align}
These equations of state involves integrals that diverge at small $\tau$, originating from floppy modes on $q_x$ and $q_y$ axes.  
To make the series expansion convergent, we assume, based on the $T=0$ solution and the assumption $\lambda\gg 1$, that (these assumptions will be verified later for the transition)
\begin{align}
	s\sim w\sim t^2 \ll \tau \ll 1 .
\end{align}
In addition, we assume that the temperature is low so that in calculating the integral from $\frac{T}{2}\ln \det \mathbf{D}$ we only need to keep terms that are singular as $\tau\to 0$, which is the leading order contribution from thermal fluctuations.  
Therefore to leading order the equations of state become
\begin{align}\label{EQ:partialF}
	\frac{\partial F}{\partial t} \Big\vert_{s,w}&= kt \left\lbrack \lambda\left( \frac{\tT}{\sqrt{\tau}}+\frac{\tau}{\lambda}+\frac{t^2}{12} \right) 
	+\frac{\tT}{\sqrt{\tau}}+s+\frac{t^2}{2}+w\right\rbrack \nonumber\\
	\frac{\partial F}{\partial s} \Big\vert_{t,w}&= k \left( \frac{2\tT}{\sqrt{\tau}}+2s+\frac{t^2}{2} +w \right)  \nonumber\\
	\frac{\partial F}{\partial w} \Big\vert_{t,s}&= k \left(  \frac{\tT}{\sqrt{\tau}}+s+\frac{t^2}{2} +w \right) ,
\end{align}
where we have defined the reduced temperature
\begin{align}
	\tT \equiv \frac{\pi T}{8 ka^2},
\end{align}
which is dimensionless.

Combining these equations we get
\begin{align}
	w=-t^2/2
\end{align}
and thus we can eliminate $w$ and Eq.\eqref{EQ:partialF} reduce to
\begin{align}
	\frac{\tT}{\sqrt{\tau}}+s &=0 \nonumber\\
	t\left\lbrack \frac{\tau}{\lambda} + \frac{\tT}{\sqrt{\tau}}+\frac{t^2}{12} \right\rbrack&=0 .
\end{align}
From these equations, it is clear that $s$ has only one solution, but $t$ has three solutions $t=0$ corresponding to the square phase, and 
\begin{align}\label{EQ:tRhom}
	t=\pm \sqrt{-\frac{12\tau}{\lambda}- \frac{\tT}{\sqrt{\tau}}} ,
\end{align}
which exist for $\tau < -\left( \tT \lambda/12\right)^{2/3}$, 
corresponding to the rhombic phase, which only exist when $\tau$ is below the transition.  (When $T=0$ this reduces to Eq.\eqref{EQ:T0EOS}). 

As discussed in the text, we then take the self-consistency approximation, replacing $\tau$ in the denominators by the corresponding fluctuation corrected shear rigidity
\begin{align}\label{EQ:rGene}
	r\equiv \frac{\partial^2 F(\stm)}{\partial t^2} =\tau +\frac{\lambda\tT}{\sqrt{r}} +\frac{\lambda t^2}{4}.
\end{align}
The correction diverges near the transition, meaning that the square phase is always locally stable, and that the transition between the square and the rhombic phase becomes a first order transition.

Therefore the phase boundary between the square and the rhombic phases are determined by the equal free energy line.  In what follows we evaluate the free energy difference $\Delta F$ between the two phases.

In the square phase, $t=0$, and the fluctuation-corrected shear rigidity satisfies the equation
\begin{align}\label{EQ:rSqua}
	r =\tau + \frac{\lambda \tT}{\sqrt{r}} .
\end{align}
In the rhombic phase, $t$ is one of the nonzero solutions to Eq.\eqref{EQ:tRhom}, and 
\begin{align}\label{EQ:rRhom}
	\frac{1}{2}r+ \frac{\lambda \tT}{\sqrt{r}} =-\tau  .
\end{align}
These two equations determine the fluctuation-corrected rigidity in the two phases (in the following discussion we use $r_0$ for the square phase and $r_1$ for the rhombic phase) for a given $\tau$.  In particular, the rhombic-phase equation~\eqref{EQ:rRhom} only has a solution for 
\begin{align}
	\tau<\kr_{c1} = -\frac{3}{2} \left(
		\tT \lambda
	\right) ^{2/3}.
\end{align}

The free energy difference then follows from the integral, as discussed in the text,
\begin{align}\label{EQ:DeltaF}
	\Delta F &= \int_{0}^{t_{\textrm{rhombic}}} \frac{\partial F}{\partial t}\Big\vert_{\partial_s F=0, \partial_w F=0}  dt \nonumber\\
	&=\int_{r_0}^{r_1} \frac{\partial F}{\partial t} \Big\vert_{\partial_s F=0, \partial_w F=0} 
	\frac{\partial t}{\partial r} \Big\vert_{\partial_s F=0, \partial_w F=0} dr .
\end{align}
To evaluate this integral, we use
\begin{align}
	\frac{\partial F}{\partial t} \Big\vert_{\partial_s F=0, \partial_w F=0}
	&= kt  \lambda\left\lbrack \frac{\tT}{\sqrt{r}}+\frac{\tau}{\lambda}+\frac{t^2}{12} \right\rbrack ,
\end{align}
and Eq.\eqref{EQ:rGene} (with $\tau\to r$ on the denominator) to eliminate $t^2$ in the above equation, and also calculate 
\begin{align}
	\frac{\partial t}{\partial r} \Big\vert_{\partial_s F=0, \partial_w F=0} =\frac{2}{t}\left\lbrack \frac{1}{\lambda} +\tT r^{-3/2}\right\rbrack .
\end{align}
The integration thus gives
\begin{align}
	\Delta F &= T \left\lbrack \frac{5\sqrt{r}}{3} + \frac{r^2}{6\lambda\tT} -\frac{\lambda \tT}{3 r} 
	- \frac{2\tau}{3\sqrt{r}} + \frac{2 r \tau}{3 \lambda\tT}\right\rbrack
	\Bigg\vert_{r_0}^{r_1} .
\end{align}
By plugging in $r_0, r_1$ we can solve for the value of $\tau$ where $\Delta F=0$.  To leading order, we get
\begin{align}
	\kr_{c2} \simeq (-\frac{3}{2}-c) \left(
		\tT \lambda
	\right) ^{2/3} ,
\end{align}
as the boundary of the first order transition between the square and the rhombic phase, where $c\simeq 0.216$, and $\kr_{c2}$ is slightly lower than $\kr_{c1}$.  This phase boundary agrees very well with our Monte Carlo simulation, as shown in the text.

\end{widetext}

\end{document}